# Observation of flat bands in gated semiconductor artificial graphene

Lingjie Du[1,2#*], Ziyu Liu[3#], Shalom J. Wind[2], Vittorio Pellegrini[4], Ken W. West[5], Saeed Fallahi[6], Loren N. Pfeiffer[5], Michael J. Manfra[6], Aron Pinczuk[2,3‡]

[1] *School of Physics, and National Laboratory of Solid State Microstructures, Nanjing University, Nanjing 210093, China*

[2] *Department of Applied Physics and Applied Mathematics, Columbia University, New York, New York 10027, USA*

[3] *Department of Physics, Columbia University, New York, New York 10027, USA*

[4] *Istituto Italiano di Tecnologia, Graphene Labs, Via Morego 30, I-16163 Genova, Italy.*

[5] *Department of Electrical Engineering, Princeton University, Princeton, New Jersey 08544, USA*

[6] *Department of Physics and Astronomy, and School of Materials Engineering, and School of Electrical and Computer Engineering, Purdue University, West Lafayette, Indiana 47907, USA*

#*L. J. D. and Z. Y. L. contributed equally to this work*

*ljdu@nju.edu.cn, ‡ap359@columbia.edu

**Flat bands near *M* points in the Brillouin zone are key features of honeycomb symmetry in artificial graphene (AG) where electrons may condense into novel correlated phases. Here we report the observation of van Hove singularity doublet of AG in GaAs quantum well transistors, which presents the evidence of flat bands in semiconductor AG. Two emerging peaks in photoluminescence spectra tuned by backgate voltages probe the singularity doublet of AG flat bands, and demonstrate their accessibility to the Fermi level. As the Fermi level crosses the doublet, the spectra display dramatic stability against electron density, indicating interplays between electron-electron interactions and honeycomb symmetry. Our results provide a new flexible platform to explore intriguing flat band physics.**

In two-dimensional electron systems (2DES), dispersionless electron bands (flat bands) present divergent density of states (DOS) [known as van-Hove singularity (vHS)]. As the Fermi level ($E_F$) crosses the vHS, electrons are usually unstable against the formation of new quantum phases such as novel superconducting states and spin or charge density waves [1-3]. Nevertheless, after extensive search, only limited electron structures have vHSs accessible for $E_F$. One famous example is Landau levels in quantum Hall effect. Another is flat bands in 'moiré' superlattices of twisted atomic layers [4,5], where superconductivity has been observed when $E_F$ overlaps a flat band of twisted-bilayer graphene [6]. In semiconductor artificial graphene (AG), pairs of flat bands with honeycomb symmetry are predicted near the Brillouin zone (BZ) *M* points [7-9]. Electron states in the semiconductor systems could be controlled by gating methods [10], giving possibilities of bringing $E_F$ to vHSs in semiconductor AG.

Semiconductor AG has electron band structures that could be tuned by honeycomb-lattice periods [9,11,12]. Linearly dispersing Dirac bands have been observed in AG based on GaAs quantum wells (QWs) [12]. Later the nanofabrication of antidots in AG provides a key element to suppress the impact of processing disorder on electrons [9]. However, probes of vHSs in the DOS of semiconductor AG that are essential to confirm the presence of flat bands have not been realized. Scanning tunneling methods offered experimental accesses to vHSs in the DOS of twisted-bilayer graphene [13] and to Dirac fermions in molecular AG [14], but are difficult to apply on semiconductor AG buried under insulating layers. Optical emission (photoluminescence, PL) could offer direct probes of the electron DOS in GaAs AG. PL spectra are from optical recombination transitions between mobile electrons in conduction bands (CB) and weakly photoexcited holes in valence bands (VB). Holes in GaAs AG have nearly dispersionless VB, so that the line shapes of PL spectra offer direct insights on the electron DOS [15-17]. The evolution of PL spectra as a function of $E_F$, enabled by gating semiconductor AG, would distinguish emerging optical characteristics.

In this Letter we report the evidence of flat bands in carrier-density-dependent PL experiments in semiconductor AG on a GaAs QW where the electron density $n_e$ is tuned by a voltage $V_b$ applied to a backgate fabricated in the device. PL spectra probe a striking emission doublet that occurs when $E_F$ crosses the flat band doublet in AG. The energy splitting of the characteristic PL doublet is well described by the DOS singularities of flat bands near $M$ points. The carrier-density dependence of the PL doublet further identifies it as the vHS doublet of flat bands. The recombination energies and line shapes of emission doublet remain constant over a wide range of $V_b$, revealing remarkable interplay between Coulomb interactions and honeycomb symmetry of electrons. The tunability of $E_F$ to access flat bands would enable explorations of novel quantum phases in nanofabricated semiconductor devices.

Figure 1(a) describes the backgated AG device structure. A high-quality 2D electron gas is confined in a 25nm single GaAs/Al$_x$Ga$_{1-x}$As QW modulation-doped with Si grown by molecular beam epitaxy [10,18]. The layer sequence and their composition have been optimized for optical measurements. An n+ Al$_x$Ga$_{1-x}$As layer serves as a backgate to tune $n_e$. As shown in Fig. 1(b), a triangular-antidot lattice with period $a = 70$ nm (equivalent honeycomb-dot-lattice period $b = 35$ nm) is patterned on the QW by means of 80 keV $e$-beam lithography followed by reactive ion etching [9,19]. The AG lattice is on a mesa fabricated by wet etching with phosphoric acid and hydrogen peroxide. Ge/Pd/Au alloy contacts are connected to the AG lattice and the backgate. The AG device is placed in an optical cryostat for measurements with a base temperature of 5 K. Figure 1(d) shows calculated AG band structures with the device parameters, where a prominent feature is a pair of flat bands around $M$ points. Figure 1(e) shows that the flat bands have vHSs in the DOS in six equivalent valleys at $M$ points.

Figure 2(a) shows $V_b$ dependence of PL spectra and Fig. 2(b) describes conduction-to-valence-band transitions active in the spectra. Holes in VB are subject to impact of disorder, because their wavefunctions have maxima under antidots [green circles in Fig. 1(b)] [9]. Three regions of $V_b$ are highlighted and their typical PL spectra are shown in Figs. 2(c)–2(e). The quasi-uniform (QU) region [red in Fig. 2(a)] is for $V_b > 0.8$ V. PL spectra in this region are dominated by a single broad peak $\Gamma_0$ at energies redshifting with increasing $V_b$ [see Fig. 2(c)]. The AG quantum limit (AL) region [green in Fig. 2(a)] is for -0.5 V $< V_b <$ 0.5 V, which is defined by the emergence of a strong PL doublet ($M_0$ and $M_1$) that is largely unchanged (see Fig. 2(d)). The low-density limit (LDL) region [yellow in Fig. 2(a)] is for $V_b < -0.8$ V, where the PL doublet finally merges into one main band ($X$) [see Fig. 2(e)].

Onset $\Gamma_0$ in the three regions is assigned to optical transitions from $c_0$ band electrons to VB

holes near the Γ point of the BZ. The broad PL band in region QU is similar to that from uniform 2DES in unpatterned QWs, whose line shape yields an accurate estimation of $E_F$ [15-17]. The evaluation of $E_F$ as a function of $V_b$ in Fig. 3 estimates population changes of AG states. It shows that at the start of region AL (0.5 V) $E_F$ is near $M_1$ singularity and moves towards $M_0$ singularity at $V_b < -0.1$ V. Remarkably, $E_F$ always stays between the singularity doublet in region AL. This implies that in region AL at 5 K optical transitions from the $M_0$ singularity make the largest contribution to PL spectra.

At $V_b = 0.5$ V, $n_e$ is estimated as $5.3 \times 10^{10}$ $cm^{-2}$ and $E_F$ (~1.9 meV) is below the AG potential 5 meV [9]. Then electrons are largely confined in the red circles [unetched area in Fig. 1(b)] that have honeycomb symmetry. In region AL the intensity step at 1518meV is continuously linked to onset $\Gamma_0$ in region QU and thus attributed to optical recombination by electrons at the Γ point (see Fig. S1 of Supplemental Material [20]). Two peaks marked as $M_0$ and $M_1$ in Fig. 2(d) are the strongest PL features in region AL. The line shapes of PL spectra in region AL show significant difference from ones in region QU and ones in the unpatterned QW (see Fig. S2 of Supplemental Material [20]). Figure 2(d) shows that the doublet is also the strongest optical emission at 10 K, which clearly reveals the presence of $M_1$ peak. The spectrum at 10 K is different from that at 5 K, which could be linked to Coulomb interaction effects as we discuss below. Under higher temperatures, electrons at Γ point are thermally populated to higher levels, giving a lower intensity of the $\Gamma_0$ line. The doublet finally disappears above 20 K.

Remarkable changes in PL occur upon entering region AL where the $\Gamma_0$ line is replaced by the strong $M_0$-$M_1$ doublet emission. This evolution of PL spectra can be interpreted as arising from changes in the DOS when 2DES evolves from a quasi-uniform status in region QU to an AG configuration in region AL where the DOS is given by honeycomb symmetry. The emergence of $M_0$-$M_1$ doublet in PL is continuous in the evolution from regions QU to AL (see Fig. S1 of Supplemental Material [20]). The energy of the $M_0$ line is 1 meV higher than that of the $\Gamma_0$ line, consistent with the situation that the singularity of the lower flat band is 1 meV higher than the CB around Γ point (see Fig. 1(e)). The energy separation between $M_0$ and $M_1$ peaks is about 0.9 meV, close to that of vHSs [relevant optical transitions are shown in Fig. 2(b)]. Thus, we could attribute the $M_0$-$M_1$ doublet to the vHS doublet of AG flat bands. The explanation of the doublet splitting is supported by the density-functional study finding that key features of AG bands are stable against electron-electron interactions [21]. The carrier-density-dependent PL experiments reveal emerging vHSs and confirm the presence of AG flat bands. Previous resonant inelastic light scattering (RILS) experiments [9,12] that probe joint DOS between AG bands suggest that the two AG bands near $M$ points are parallel, but cannot encode the energy dispersion of each single flat band (see Fig. S3 of Supplemental Material [20]).

The honeycomb symmetry of AG bands is linked to the constancy of doublet energies. As shown in Figs. 2(a) and 4(a), the PL doublet energies do not change in region AL, which could be interpreted in terms of the symmetry of the DOS singularity at $M$ points of the BZ where there are six equivalent valleys sharing $n_e$. Figure 3 shows that $n_e$ in region AL is typically about $4 \times 10^{10}$ $cm^{-2}$. We estimate that $n_e$ in each valley of the $M_0$ singularity is tuned down by $V_b$ from $2.3 \times 10^9$ $cm^{-2}$ when states of the $M_0$ singularity are fully populated, to nearly full depletion when $E_F$ is below the singularity. This is a relatively small density variation that would not change the optical recombination energy of electrons in GaAs QWs [22,23].

The discussions above demonstrate that the energy properties of PL doublet can be captured by the single-particle picture based on honeycomb symmetry. Nevertheless, detailed PL spectra behaviors cannot be understood by single-particle physics. The line shapes of PL spectra in region AL cannot be described by calculations based on AG DOS. The intensity ratio of the $M_0$ line over the $M_1$ line in experiments is lower than those in single-particle calculations (see Fig. S4 of Supplemental Material [20]). Moreover, as shown in Fig. 4(a), the intensity of $M_0$ and $M_1$ transitions is fixed for $V_b$ > -0.1 V, and their ratio is fixed across region AL. In a single-particle picture, this behavior indicates that the population ratio between $M_0$ and $M_1$ singularities is fixed, which is unlikely to happen because $E_F$ sweeps from one singularity to another (see Fig. S5 of Supplemental Material [20]).

Many-body interactions, including exchange interactions in the flat band doublet, would play an important role in understanding detailed PL spectra behaviors. In region AL, $E_F$ stays between the singularity doublet (see Fig. 3), and the AG states are thus similar to those reported in Ref. [9], where large exchange interactions between AG bands (the exchange energy is about 0.6 meV ≈ 8 K) were reported. We could get some insight about this physics process from the Coulomb coupling between Fermi edge singularity and higher sub-band excitons [24-27]. Here the lower flat band is heavily populated while the upper flat band is weakly populated by thermal excitations and basically empty, thus the $M_1$ line intensity should be low under single-particle physics. However, the exchange energy between electrons in two flat bands is comparable with the splitting of flat bands, and Coulomb coupling would provide extra scattering channels from the lower flat band below $E_F$ to the upper flat band. The strong coupling between the flat band doublet would contribute to a considerable luminescence intensity to $M_1$ singularity, and thus strengthens the ratio of the $M_1$ line intensity over the $M_0$ line intensity.

On the other hand, due to honeycomb symmetry, each valley has a small $n_e$. Under the small $n_e$ born with the AG device, the Coulomb-interaction terms between the flat band doublet that are dominated by exchange coupling processes should have weak dependence on $n_e$ and $V_b$ [22,23]. Therefore, the doublet intensity is stable as long as the Coulomb coupling dominantly modifies the luminescence intensity. The striking stability of PL spectral line shapes in region AL suggests many-body interactions interplayed with symmetry of AG electrons. This can be confirmed by PL spectra under high temperatures. As shown in Fig. 2(d), for a higher temperature 10 K exceeding the exchange energy, the PL spectral line shape becomes unstable and the population ratio between $M_0$ and $M_1$ singularities changes. The larger $M_1$ line intensity at 10 K would be attributed to the major effect of excitons from thermally elevated electrons into the upper flat band [24]. A detailed line shape analysis should consider evolution of localized states and is out of the scope of this Letter.

The Coulomb coupling would be modulated as $E_F$ crosses the lower flat band for $V_b$ < -0.1 V (see Fig. 3). Figure 4(a) shows that the PL intensity starts to drop for $V_b$ < -0.1 V, indicating the beginning of depopulation of $M_0$ singularity (see insets of Fig. 4(a)). At $V_b$ < -0.9 V, $M_0$ transition quickly collapses and the doublet evolves into one broad band labeled $X$ [Fig. 4(b)]. The contrasting line shapes in regions LDL and AL indicate different underlying processes. At $V_b$ = -1.1 V, $E_F$ is well below $M_0$ singularity and the population of $M_0$-$M_1$ doublet is greatly reduced, so optical transitions from these states can be considered as excitonic. Instead of a doublet, band $X$ in Fig. 4(b) indicates the mixture of excitonic transitions from each singularity by strong coupling between the doublet [9]. The attribution of band $X$ to excitons is consistent with significant redshifts of its PL energies in region LDL [see Figs. 2(a) and 4(b)]. The operation of $V_b$ to reduce $n_e$ results in an increase of the electric field at the QW. Then, the redshift could be understood in terms of a quantum-confined-Stark effect of excitons [28] that brings redshifts of about 0.5 meV for electric field changes of $5 \times 10^3$ V/$cm$. We note that the appearance of excitonic transitions above $E_F$ is nontrivial and implies couplings between the Fermi sea and higher AG bands [24, 25].

To summarize, flat bands are observed in carrier-density-dependent PL experiments in a backgated semiconductor AG device. PL spectra show striking dependence on carrier densities tuned by $V_b$, marked by three regions with contrasting spectral line shapes, and proves that the flat bands of semiconductor AG are accessible by $E_F$. Under appropriate flat bands population, Coulomb interactions within the flat band doublet are observed and take a critical role in stability of PL spectra. Several key features of semiconductor AG make it an excellent platform for further studies. For example, transport measurements could investigate many-body physics in the embedded flat bands, e.g., unconventional superconductivity [2,3,29]. Under strong magnetic fields [30], correlated phases such as Mott-Hubbard bands observed in honeycomb lattices [31] could promise novel collective behaviors. In general, compared with 2D moiré heterostructures, AG can be employed as a seminal quantum simulator of electron correlation operating in a rarely studied density regime.

**Acknowledgement**

The work at Columbia University was supported by the National Science Foundation, Division of Materials Research under award No. DMR-1306976 and by grant No. DE-SC0010695 funded by the U.S. Department of Energy Office of Science, Division of Materials Sciences and Engineering. Work at Nanjing University was supported by the Fundamental Research Funds for the Central Universities (Grant No. 14380146) and National Natural Science Foundation of China (Grant No. 12074177). This research is funded in part by the Gordon and Betty Moore Foundation's EPiQS Initiative, Grant No. GBMF9615 to L. N. P., and by the National Science Foundation MRSEC grant No. DMR 1420541. Work at Purdue University was supported by the U.S. Department of Energy, Office of Science, Basic Energy Sciences, under Award No. DE-SC0006671.

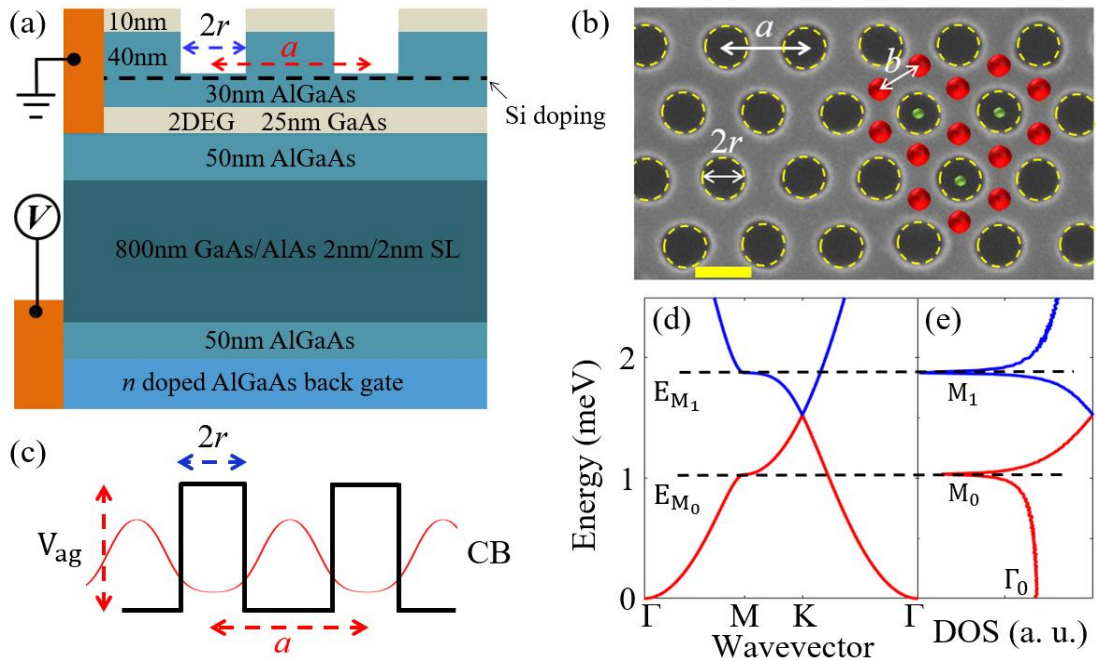

Fig. 1 (color online). (a) Cross-section view of the AG device showing the heterostructure layer sequence with triangular-antidot lattices imprinted on a GaAs QW. Dimensions are not to scale. The antidot radius is $r$ and the period is $a$. SL is for superlattice. (b) Scanning electron microscopy micrographs of AG lattices with $a = 70$ nm and $r = 20$ nm. The dashed circles mark antidots. The variation of $r$ is below 5 nm. The scale bar is 50 nm. Red large dots indicate maximum positions of electron wavefunctions. Green small dots indicate positions of photoexcited holes. Holes are in a triangular lattice with a large effective mass, resulting in nearly dispersionless VB. (c) Muffin-tin AG potential and wavefunctions in the single-particle approximation for electrons. Panels (d) and (e) show the two lowest AG bands and DOS with the parameters in (b). In (e), the $\Gamma$ point onset, the singularity from a lower flat band and the singularity from an upper flat band are marked as $\Gamma_0$, $M_0$, and $M_1$, respectively. $E_{M_1}$ and $E_{M_0}$ mark energies of the upper and lower flat band.

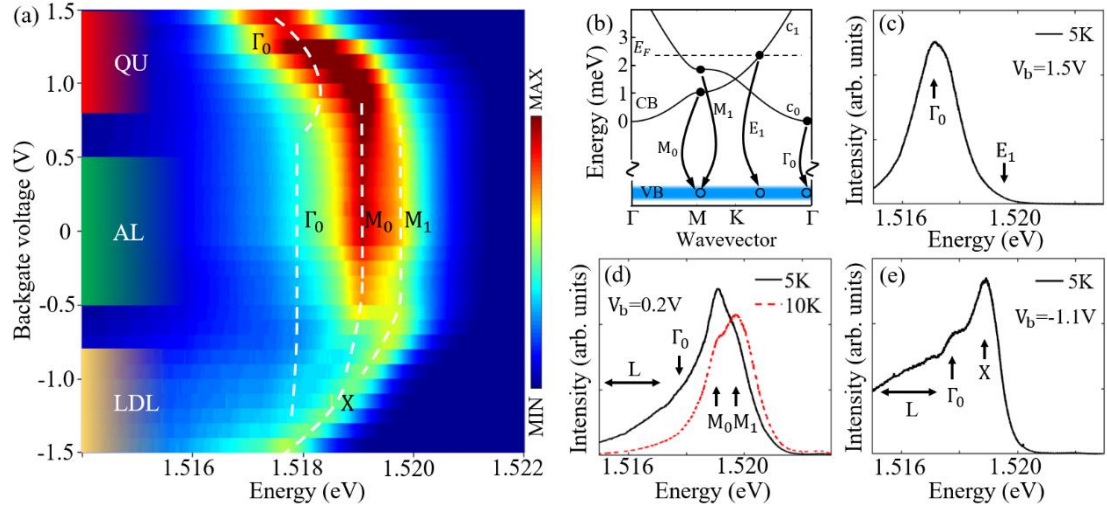

Fig. 2 (color online). (a) PL spectra as a function of $V_b$ at 5 K. PL spectra were excited by tunable emissions from a Ti:sapphire laser focused to a spot of ~100 μm onto the AG device. Incident power was 3-10 μW. Dashed lines mark PL peaks as $\Gamma_0$, $M_0$, $M_1$, and $X$. The color code is linear with intensity. (b) Optical transitions in PL peaks between electron (dot) and hole (circle) states. $E_1$ represents the energy of transitions at $E_F$. The two lowest AG CBs are marked as $c_0$ and $c_1$. (c) PL trace for $V_b = 1.5$ V at 5 K. The difference between $\Gamma_0$ and $E_1$ yields a determination of $E_F$. (d) The black (dotted red) line represents PL trace under $V_b = 0.2$ V at 5 K (10 K). $L$ indicates the transitions from localized states. (e) PL trace under $V_b = -1.1$ V at 5 K.

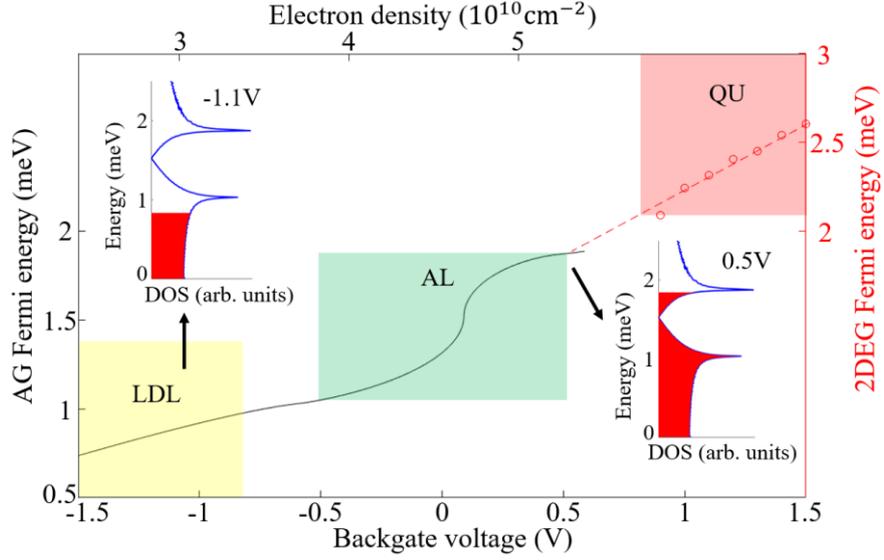

Fig. 3. (color online). $E_F$ as a function of $V_b$ ($n_e$). Red open circles and the red dashed line represent $E_F$ determined from PL spectra and their linear extension. In region QU, $E_F$ is transferred to $n_e$ from difference between $\Gamma_0$ and $E_1$ as shown in Fig. 2(b) since electrons are quasi-two-dimensional. Because $V_b$ tunes $n_e$ linearly, it determines other $n_e$ marked in the top axis. The black line represents the calculated $E_F$. Three regions in red, green and yellow are those defined in Fig. 2(a). Insets: schematic representations of AG band populations at $V_b$ = -1.1 V and 0.5 V.

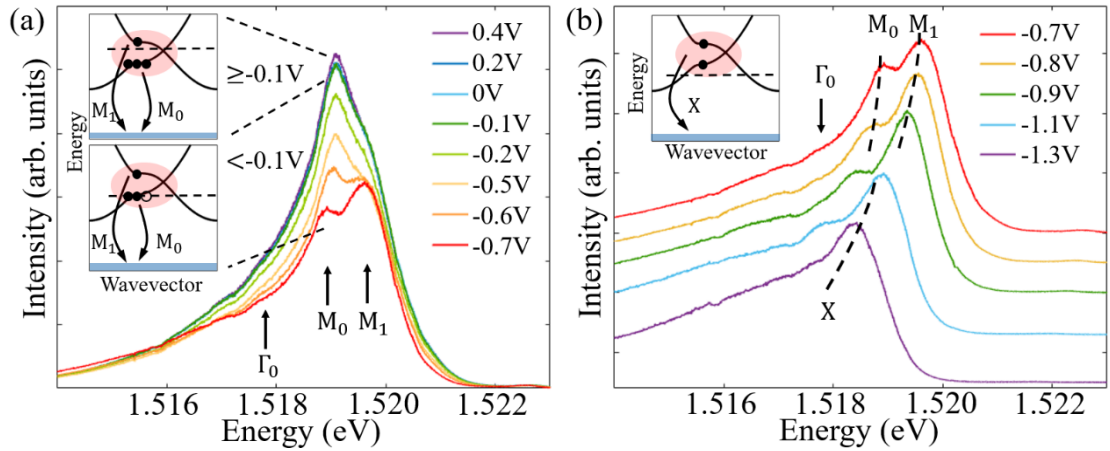

Fig. 4. (color online). (a) PL spectra in region AL. (b) PL spectra in the transition from AL to LDL regions. Insets: optical transitions at different voltage ranges. Filled and empty dots represent population changes in each flat band and the dashed red line marks $E_F$. The elliptical background denotes coupling between transitions. PL spectra of an unpatterned QW sharing the same contacts with the AG device show that its $E_F$ is linear with $V_b$. In (b) at low $n_e$ two excitonic transitions mix with each other and give a coupled excitonic transition (X). The large peak width comparable to the doublet energy splitting in (a) and (b) can be attributed to both thermal distribution of holes and electron-electron interactions. The lifetime broadening of electron states due to disorder is small (suggested by narrow intersubband RILS peaks), and the peak width does not represent the sharpness of AG DOS.

# Supplementary Material

# Observation of flat bands in gated semiconductor artificial graphene

Lingjie Du[1,2#*], Ziyu Liu[3#], Shalom J. Wind[2], Vittorio Pellegrini[4], Ken W. West[5], Saeed Fallahi[6], Loren N. Pfeiffer[5], Michael J. Manfra[6], Aron Pinczuk[2,3‡]

[1] School of Physics, and National Laboratory of Solid State Microstructures, Nanjing University, Nanjing 210093, China

[2] Department of Applied Physics and Applied Mathematics, Columbia University, New York, New York 10027, USA

[3] Department of Physics, Columbia University, New York, New York 10027, USA

[4] Istituto Italiano di Tecnologia, Graphene Labs, Via Morego 30, I-16163 Genova, Italy.

[5] Department of Electrical Engineering, Princeton University, Princeton, New Jersey 08544, USA

[6] Department of Physics and Astronomy, and School of Materials Engineering, and School of Electrical and Computer Engineering, Purdue University, IN, 47907, USA;

[#] L. J. D. and Z. Y. L. contributed equally to this work

*ljdu@nju.edu.cn, ‡ap359@columbia.edu

## 1. Formation of artificial graphene potential

The formation of artificial graphene (AG) potential is revealed by the continuous evolution of photoluminescence (PL) spectra from Quasi-Uniform (QU) region to AG Quantum Limit (AL) region. As shown in Fig. S1, in the transition from QU region to AL region ($0.5\ V < V_b < 1.2\ V$), the intensity step in the AL region at 1518meV is continuously linked to the broad PL maximum $\Gamma_0$ in the QU region and is thus attributed to optical recombination by charge carriers at the $\Gamma$-point of the BZ. This transition from QU region to AL region also shows that the $M_0$ and $M_1$ lines start to appear in the spectra. The emergence of $M_0$ - $M_1$ doublet is the direct evidence of the van-Hove singularities in the density of states (DOS) of semiconductor AG.

PL spectra that directly probe the electron DOS in GaAs AG are in stark contrast to the ones in the QU region of the AG device and ones in the unpatterned quantum well (Fig. S2). Both energies and lineshapes of the spectra show significant differences due to the formation of AG potential.

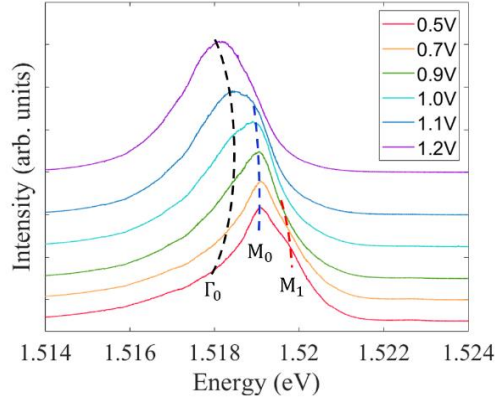

Figure S1. Evolution of photoluminescence spectra from Quasi-Uniform region to AG Quantum Limit region. The $M_0$ line starts to appear at 1.1 V where the $\Gamma_0$ line is still dominant. At 1.0 V, the $M_0$ line has become dominant over the $\Gamma_0$ line. The $M_1$ line starts to appear at 0.9 V.

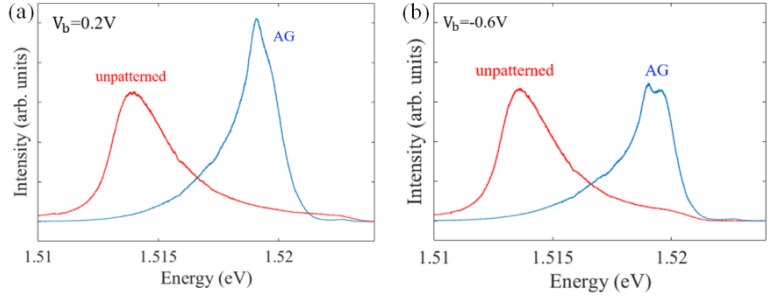

Figure S2. Comparison between photoluminescence spectra for the unpatterned quantum well and the artificial graphene at back-gate voltage of (a) 0.2V and (b) -0.6V.

## 2. Comparison between photoluminescence and resonant inelastic scattering measurements

PL (probing density of states) and resonant inelastic light scattering (RILS) (probing joint density of states) are two different technologies, handling distinct physics problems. As shown in Fig. S3, for parallel non-flat bands in Fig. S3(a) and parallel flat bands in Fig. S3(d), RILS spectra give the same results (see Figs. S3(b) and S3(e)), so RILS can only confirms that the two bands are parallel, but cannot encode the energy band dispersion of single bands. In contrast, for parallel non-flat bands in Fig. S3(a) and parallel flat bands in Fig. S3(d), PL spectra give distinct results (see Figs. S3(c) and S3(f)) and reveal the energy band dispersion of single flat bands by probing singularities of the flat bands. In the current work, we observe van-Hove singularity doublet of M-point flat bands by charge carrier density dependent PL, which presents the evidence of the flat bands for AG semiconductor.

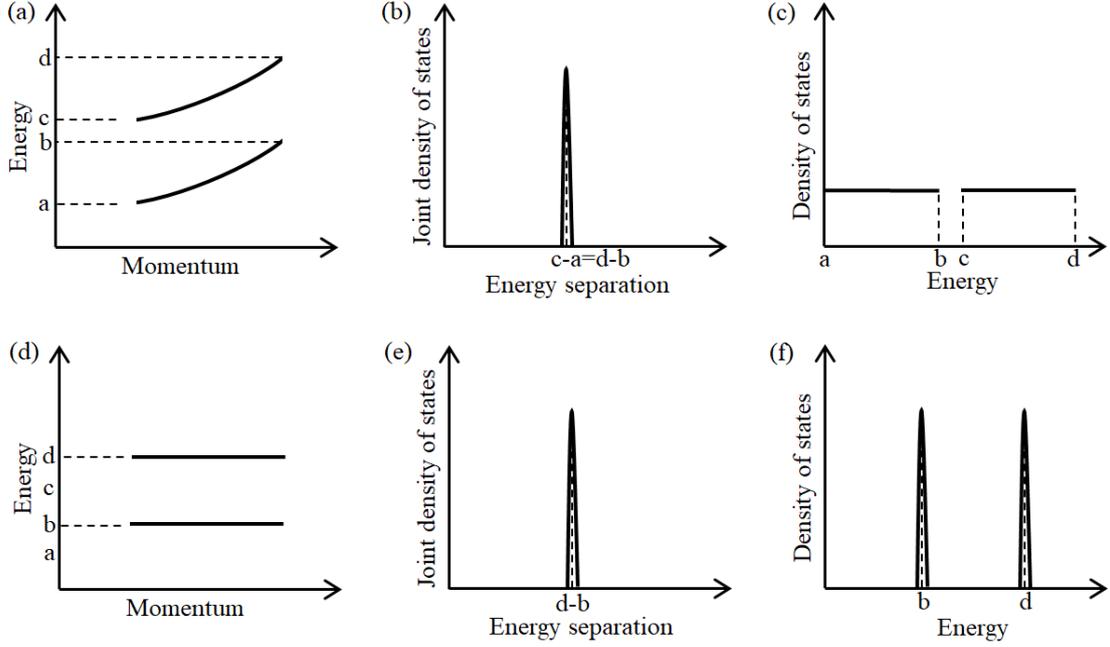

Figure S3. Comparison between resonant inelastic light scattering and photoluminescence measurements. (a) shows two parallel not-flat bands. The transition intensity between the two bands, which is measured by resonant inelastic light scattering, is shown as joint density of states in (b). The density of states of the bands in (a), which is measured by photoluminescence, is shown in (c). Although there is a peak for the transition intensity between the two bands, there is no peak in the density of states because the bands are not flat. On the other hand, (d) shows two parallel flat bands. The transition intensity between the two bands, which is measured by resonant inelastic light scattering, is shown as joint density of states in (e). The density of states of the bands in (d), which is measured by photoluminescence, is shown in (f). We can see that there are two peaks in the density of states for the two flat bands. Therefore, the measurement of the density of states, in our case photoluminescence measurement, is the way to distinguish the flat bands.

## 3. Failure of single-particle physics in explanations of optical spectra

Detailed PL spectra of AG could not be understood by single-particle physics. According to a single-particle picture, we calculate lineshapes of the PL spectra in AL region based on the product of the Gaussian broadened DOS of AG and Fermi-Dirac distribution. As shown in Fig. S4, in AL region (at back-gate voltage of 0.2 V, corresponding Fermi energy 1.7 meV) the ratio of the $M_0$ line intensity over the $M_1$ line intensity in experiments is higher than those in the single-particle calculations. Tuning the full width at half maximum (FWHM) of Gaussian broadening cannot account for this discrepancy. In Fig. S5, we compare the carrier density dependence of the calculated and measured PL spectra (-0.1 V < $V_b$ < 0.4 V, corresponding range of Fermi energies 1.1 meV < $E_F$ < 1.9 meV). In the single-particle picture, the population ratio between $M_0$ and $M_1$ singularities changes considerably as the Fermi level sweeps from $M_0$ singularity to the other (Figs. S5(a) and S5(b)). However, the measured spectra present striking stability and the single-particle calculations can't describe the experimental results (Fig. S5(c)).

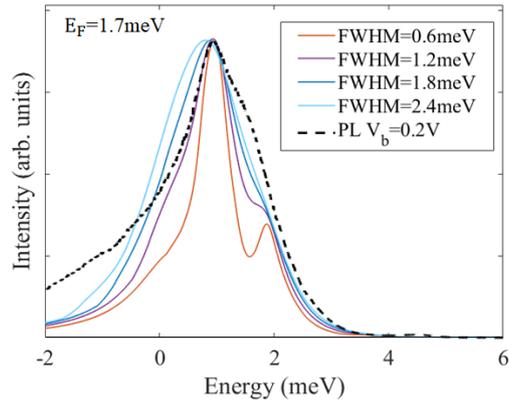

Figure S4. Comparison between calculated traces with different full widths at half maximum (FWHM) at the Fermi energy of 1.7 meV. The dashed line is the measured photoluminescence spectrum at back-gate voltage of 0.2 V.

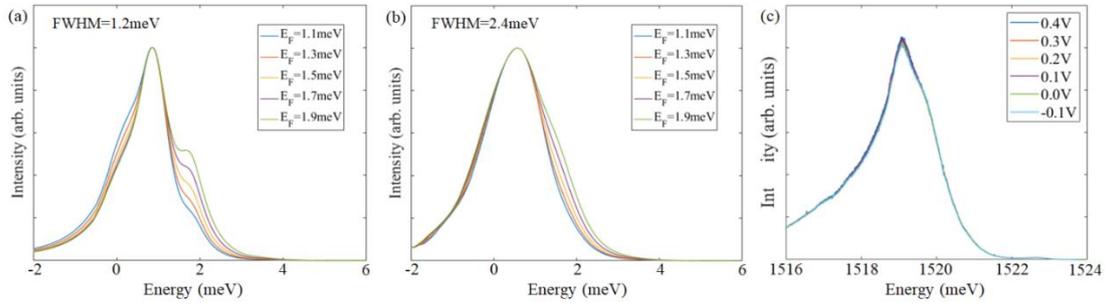

Figure S5. Comparison between measured photoluminescence spectra and calculations based on the single-particle picture. (a) and (b) Calculated traces at different Fermi energies with a full width at half maximum (FWHM) of 1.2 meV and 2.4 meV, respectively. (c) Measured photoluminescence spectra at different back-gate voltages corresponding to the Fermi energies in (a).